\documentclass[prd,aps,showpacs,nofootinbib,twocolumn]{revtex4}
\usepackage{amssymb}
\usepackage{graphicx}
\usepackage{amsmath}
\usepackage{latexsym}
\usepackage{rotating}
\usepackage{amstext}
\usepackage{tabularx}
\usepackage{subfigure}
\usepackage{amssymb}
\usepackage[all]{xy}
\usepackage{times}
\usepackage{bbm}
\usepackage{epsf}
\usepackage{pstricks}
\usepackage{amsfonts}

\begin{document}

\title{Aharonov-Bohm-Casher Problem with a nonminimal Lorentz-violating
coupling}
\author{H. Belich $^{a,b,d}$, E. O. Silva$^{b}$, M. M. Ferreira Jr.$^{c}$,
M. T. D. Orlando$^{a,d}$}
\affiliation{$^{a}${\small {Universidade Federal do Esp\'{\i}rito Santo, Departamento de F
\'{\i}sica e Qu\'{\i}mica, Av. Fernando Ferrari, 514, Goiabeiras, 29060-900,
Vit\'{o}ria, ES, Brazil}}}
\affiliation{$^{b}${\small {International Institute of Physics, Universidade Federal do
Rio Grande do Norte, Av. Odilon Gomes de Lima, 1722, 59078-400, Natal, RN,
Brazil}}}
\affiliation{$^{c}${\small {Departamento de F\'{\i}sica, Universidade Federal Maranh\~{a}
o, Campus Universit\'{a}rio do Bacanga, 65085-580, S\~{a}o Lu\'{\i}s, MA,
Brazil}}}
\affiliation{$^{d}${\small {Grupo de F\'{\i}sica Te\'{o}rica Jos\'{e} Leite Lopes, C.P.
91933, 25685-970, Petr\'{o}polis, RJ, Brazil}}}
\email{belichjr@gmail.com, edilbertoo@gmail.com, manojr07@ibest.com.br,
orlando@cce.ufes.br}

\begin{abstract}
The Aharonov-Bohm-Casher problem is examined for a charged particle describing a circular path in presence of a Lorentz-violating background nonminimally coupled to a spinor and a gauge field. It were evaluated the particle eigenenergies, showing that the LV background is able to lift the original degenerescence in the absence of magnetic field and even for a neutral particle. The Aharonov-Casher phase is used to impose an upper bound on the background magnitude. A similar analysis is accomplished in a space endowed with a topological defect, revealing that both the disclination
parameter and the LV background are able to modify the particle eigenenergies. We also analyze the particular case where the particle interacts harmonically with the topological defect and the LV background, with similar results.
\end{abstract}

\maketitle

\section{Introduction}

The Standard Model Extension (SME) \cite{Colladay, Samuel} is the natural framework for studying properties of physical systems with Lorentz-violation once it includes Lorentz-violating terms in all sectors of the minimal standard model, including gravitation \cite{Gravity}. The Lorentz violation terms are generated as vacuum expectation values of tensors defined in a high energy scale. The SME is a theoretical framework which has inspired a great deal of investigations in this theme in recent years. Such works encompass several distinct proposes involving: fermion systems \cite{fermion}, CPT- probing experiments \cite{CPT}, the electromagnetic CPT- and Lorentz-odd term \cite{Adam,photons1}, the nineteen electromagnetic CPT-even and Lorentz-odd coefficients \cite{photons2,Cherenkov2}, topological defects \cite{Defects}, topological phases \cite{Phase1,Phase2}, cosmic rays \cite{CosmicRay}, and other relevant aspects \cite{General2}. These many contributions have elucidated the effects induced by Lorentz violation and serve to set up stringent upper bounds on the Lorentz-violating (LV) coefficients.

Further investigation on the influence of Lorentz symmetry violation on fermion systems has been accomplished in Ref. \cite{Phase1}, where a Lorentz-violating background four-vector, $\mathrm{v}^{\mu}$, was taken in a nonminimal coupling with the gauge field and the fermion spinor ($\Psi$) in the Dirac equation, $(i\gamma^{\mu}D_{\mu}-m)\Psi=0.$ In this case, the
nonminimal covariant derivative was defined as
\begin{equation}
D_{\mu}=\partial_{\mu}+eA_{\mu}+i\frac{g}{2}\epsilon_{\mu\lambda\alpha\beta}\mathrm{v}^{\lambda}F^{\alpha\beta},  \label{cov}
\end{equation}
where $g$\ is the constant that measures the strength of this coupling, and $\gamma^{\mu}=(\gamma_{0},\gamma^{i})$ are the usual gamma matrices. Here, it is supposed that the Lorentz violating background $\mathrm{v}^{\mu}$ appears as the vacuum expectation value of a tensor quantity generated in a spontaneous breaking of symmetry taken place in a fundamental theory defined at a higher energy scale (a consequence of an anisotropic vacuum at Planck scale). The product $g\mathrm{v}^{\mu}$ has mass dimension equal to $-1.$The analysis of the nonrelativistic regime of this theory revealed that such a coupling is able to induce Aharonov-Casher (AC) \cite{Casher} geometrical phases on the wave function of an electron interacting with the gauge field. Moreover, the background induces a magnetic moment for neutral particles, which explains the generation of AC phase for a chargeless particle (deprived from magnetic moment) in this context. Next, still in connection with this particular nonminimal coupling, it was found out that particles and antiparticles may develop opposite AC phases \cite{Phase2}. It is worthy stressing that the standard AC phase is currently interpreted as due to a Lorentz change in the observer frame. In our proposal, namely, in a situation where Lorentz symmetry is violated, it rather emerges as a phase whose origin is ascribed to the presence of a privileged direction in the space-time, set up by the fixed background. Since in this kind of model Lorentz invariance in the particle frame is broken, the AC effect could not any longer be obtained by a suitable Lorentz change in the observer frame. An additional study involving such a coupling was developed in Ref. \cite{Nonmini}, where it were carried out the corrections yielded on the hydrogen spectrum, allowing the establishment of some upper bounds on the background magnitude.

The study of systems with topological phases was initiated with the pioneering work of Aharonov-Bohm \cite{ab}, being followed by a series of other relevant works discussing the conditions and properties of such effect \cite{ab2}. The discovery of Aharonov-Casher effect \cite{Casher} revealed that the generation of geometrical phases could be triggered by an electric field, motivating a great number of interesting investigations \cite{Casher2}. The first experimental confirmations of the AC effect \cite{cimmino} reported phase measurements as small as $10^{-3}$ rad, with precision in the
order of $10^{-4}$ rad.

The influence of geometrical phases on condensed matter mesoscopic systems started to be studied on one-dimensional (1D) conducting rings since the early 90\'{}s. This issue was initially addressed under the action of an external
magnetic field, in which the AB effect is responsible by the Altshuler-Aronov-Spivak (AAS) conductance oscillations \cite{AAS}. In 1992, it was proposed that the AC effect could engender conductance oscillations in semiconducting 1D rings \cite{Mathur} as well as the AB effect yields the AAS oscillations. Further developments revealed that the Rashba spin-orbit interaction \cite{SOI} and the AC effect can be used to control the conductance on 1D mesoscopic rings \cite{Nitta}, with experimental demonstrations \cite{Konig}. This issue is also discussed in some books \cite{Ihn}.

Another point of interest concerns the influence of topology (inherent to topological defects) on the usual AB and AC phases. These defects are characterized by a spacetime metric with a Riemann-Christoffel curvature tensor which is null everywhere except on the defects. Here, this curvature conical singularity topological defect is identified by a cosmic string. The cosmic string is used to study the gravitational analogue of the bound-state AB effect. Instead of a magnetic flux is considered the curvature flux provided by a cosmic string \cite{strings}. This object, geometrically, corresponds to a Minkowski space-time with a conical singularity, that is, a line element given by
\begin{equation}
ds^{2}=-c^{2}dt^{2}+dr^{2}+\alpha^{2}r^{2}d\phi^{2}+dz^{2},\label{metrica}
\end{equation}
where $r\geq0$ and $0\leq\phi\leq2\pi$. This metric has a cone-like singularity at $r=0$. The parameter $\alpha$, which effectively introduces an angular deficit $2\pi\left( 1-\alpha\right) $\ in the Minkowski geometry, is related to the linear mass density $\mu$\ of the string by $\alpha =1-4G\mu/c^{2}$, where $G$\ is the gravitational constant and $c$\ is the
speed of light.

The curvature tensor of the defect is given by
\begin{equation}
R_{12}^{12}=R_{1}^{1}=R_{2}^{2}=2\pi\left( \frac{\alpha-1}{\alpha}\right)\delta\left( \mathbf{r}\right),\label{tensor}
\end{equation}

This tensor characterizes a two-dimensional conical singularity labeled by the $\delta$-function. This defect is symmetrical in the z-axis, what implies it be a linear defect. In quantum mechanics, when dealing with problems involving singularities, we can not impose that the wave function of the particle is regular in all space including the singular point \cite{hagen1}. Just as happens for AB effect, a singularity is also regarded as a topological effect. Its physical properties can only be studied at the quantum level. The reason is because the particle does not have access to the core of such singularity. Although the particle does not have access to this region, its wave function and energy spectrum are influenced by the core.

A useful method to study the behavior of wave functions for a particle bound in the region inaccessible (core) is the self-adjoint extensions \cite{reed}. The problem dealing about self-adjoint extensions of operators appears in various contexts in quantum physics \cite{selfs}. In such works it was found the family of the self-adjoint extensions. Although the authors had discussed the role of the extensions parameter we can not see them expressed in terms of the physics of the problems. One way to self-select the parameter extension is to modulate the problem by boundary conditions, as proposed by Kay and Studer \cite{kay}. In fact, the physics of the problem must selects such extension parameter \cite{self}. Here, we will not address the problem via the method of self-adjoint extension, because our main goal is to analyze some quantum properties of a particle moving in a plane, having excluded the origin region.

In this work, we analyze the effects of a LV background vector, nonminimally coupled to the gauge and fermion fields, on the Aharonov-Bohm-Casher (ABC) problem of a charged particle in a circular motion in a plane, in the presence of magnetic and electric fields. We also study the quantum dynamics of a charged particle in the presence of a conical singularity and the
particular case where the particle interacts harmonically with a conical singularity. This paper is organized as follows. In Sec. \ref{sec2} we formally introduce the nonminimal coupling and write the Hamiltonian of the system. In Sec. \ref{sec3} we study the ABC problem in the context of the LV nonminimal coupling. We determine the expressions for the energy eigenvalues, the geometric phase and discuss the role played by the nonminimal background on the eigenenergies and the AC phase. In Sec. \ref{sec4} we analyze the topological ABC problem in the presence of Lorentz violation. We study the effects of topology and the LV background field $v$ on the energy and geometric phase, comparing the results with the ones of the original problem. In Sec. \ref{sec5} the topological ABC interacting with a two-dimensional harmonic oscillator is studied. Expressions for the
energy eigenfunctions and eigenvalues are found. In Sec. \ref{sec6} a brief conclusion is given.

\section{The non-minimal coupling}

\label{sec2}

The nonminimal coupling of fermions to a gauge field and a LV background four-vector was proposed and analyzed in Refs. \cite{Phase1,Phase2}. In these works, after assessing the nonrelativistic regime, one has identified a generalized canonical momentum,
\begin{equation}
\mathbf{\Pi }=\left( \mathbf{p}-q\mathbf{A}+g\mathrm{v}_{0}\mathbf{B}-g \mathbf{v}\times \mathbf{E}\right) .
\end{equation}
where the background is $\mathrm{v}^{\mu }=(\mathrm{v}_{0},\mathbf{v})$. This result allows to introduce this nonminimal coupling in an operational way, that is, just redefining the vector potential and the corresponding magnetic field as indicated below:
\begin{eqnarray}
\overset{\thicksim }{\mathbf{A}} &=&\mathbf{A}-\frac{g}{q}\mathrm{v}_{0}\mathbf{B}+\frac{g}{q}\left(\mathbf{v}\times \mathbf{E}\right),\label{A1}
\\
\overset{\thicksim }{\mathbf{B}} &=&\boldsymbol{\nabla }\times \mathbf{A}-\frac{g}{q}\mathrm{v}_{0}\boldsymbol{\nabla }\times \mathbf{B}+\frac{g}{q}\boldsymbol{\nabla }\times \left( \mathbf{v}\times \mathbf{E}\right),
\label{A2}
\end{eqnarray}
so that the generalized canonical momentum recoveries the usual form, $\mathbf{\Pi}=(\mathbf{p}-q\overset{\thicksim}{\mathbf{A}})$. Taking these prescriptions into account, we obtain the following interaction Hamiltonian density \textbf{(}written in natural units\textbf{)}:
\begin{eqnarray}
H &=&\frac{1}{2m}\mathbf{\Pi }^{2}-\boldsymbol{\mu }\cdot \mathbf{B}+\frac{1}{2m}g\text{v}_{0}\boldsymbol{\sigma }\cdot (\boldsymbol{\nabla }\times \mathbf{B})+  \notag \\[-0.2mm] && \\[-0.2mm]
&+&\frac{g}{2m}\boldsymbol{\sigma }\cdot \boldsymbol{\nabla }\times \left( \mathbf{v}\times \mathbf{E}\right) +qU, \notag  \label{H1}
\end{eqnarray}
where $q,\mu =\frac{q}{2m}\sigma $ $\ $are the charge and the magnetic momentum dipole of the particle, while $U$ is the electrostatic potential due to linear charge distribution. For a situation where the magnetic field is constant and $\boldsymbol{\nabla }\times \left( \mathbf{v}\times \mathbf{E}\right) =0,$ the Hamiltonian is
\begin{equation}
H=\frac{1}{2m}\left[ \mathbf{p}-q\mathbf{A}+g\mathrm{v}_{0}\mathbf{B}-g \mathbf{v}\times \mathbf{E}\right] ^{2}-\boldsymbol{\mu }\cdot \overset{\thicksim }{\mathbf{B}}+qU.\label{H2}
\end{equation}
The nonminimal coupling of the background with the electromagnetic field generates a magnetic dipole moment $g\mathbf{v}$ \cite{Phase1} even for a chargeless particle. In the next section, we will study the dynamics of a charged particle describing a circular path and governed by this Hamiltonian.

\section{The Aharanov-Bohm-Casher problem in the context of the LV nominimal coupling}\label{sec3}

In this section, we reassess the motion of a particle in a circular path of mesoscopic dimension, subject to the action of an orthogonal magnetic fields and a radial electric field. This system is ruled by the Schr\"{o}dinger
equation, once here we are disregarding the spin degree of freedom. This kind of system may have connection with the physics of 1D mesoscopic rings analyzed in Refs. \cite{Mathur,SOI,Nitta}.

The magnetic flux tube is the confined into a solenoid orthogonal to the plane, being specified by
\begin{equation}
\mathbf{B}=\frac{\Phi _{(B)}}{r}\delta \left( r\right) \delta \left( \phi \right) \hat{z},\hspace{0.5cm}\mathbf{A}=\frac{\Phi _{(B)}}{2\pi r}\hat{\phi}, \label{ba}
\end{equation}
where $\Phi _{(B)}$ is the magnetic flux and the vector potential $\mathbf{A}$ is aligned in the direction $\hat{\phi}$. The magnetic field is confined along the z-axis. Moreover, let us admit that the solenoid has a linear charge density $\lambda $ distributed uniformly along its length, which engenders the following electric field:
\begin{equation}
\mathbf{E}=\frac{\Phi _{(E)}}{2\pi r}\hat{r}.  \label{ce}
\end{equation}
In order to examine the role played by the nonminimal coupling (\ref{cov}) on this system, we take the Hamiltonian (\ref{H2}) with the fields configurations (\ref{ba}) and (\ref{ce}). The background field $\mathbf{v}$ is chosen parallel to the magnetic field $\mathbf{B}$ so that the term $\left( \mathbf{v}\times \mathbf{E}\right)$ will contribute to the geometric phase. The particle is allowed to move in a region where the magnetic field, $\mathbf{B}$, is null. So the Hamiltonian (\ref{H2}) takes the form,
\begin{eqnarray}
H &=&\frac{1}{2m}\Big[\left( \mathbf{p}-q\mathbf{A}\right) ^{2}-2g\left(\mathbf{v}\times \mathbf{E}\right) \cdot \left( \mathbf{p}-q\mathbf{A}\right) +  \notag \\[-0.2mm]
\\[-0.2mm]
&+&g^{2}\left(\mathbf{v}\times \mathbf{E}\right) ^{2}\Big]+qU.\notag
\end{eqnarray}
As well-known, the kinetic part of Hamiltonian, $\left(\mathbf{p}-q\mathbf{A}\right) ^{2}/2m$, in polar coordinates, is written as
\begin{equation}
\frac{1}{2mr_{0}^{2}}\left[ \frac{\Phi }{\Phi _{0}}+i\frac{\partial}{\partial \phi }\right]^{2}-\frac{1}{2m}\left[\frac{\partial ^{2}}{\partial
r^{2}}+\frac{1}{r}\frac{\partial }{\partial r}\right],
\end{equation}
and we have used $\hbar = 1,\boldsymbol{\nabla }\cdot \mathbf{A}=0$, $\Phi_{0}=2\pi /q.$ Further, if the particle is constrained to move in a path with constant radius, $r=r_{0}$ (with $\theta =\pi /2$), the wave function
will depend only on the azimuthal angle $\phi $, and the Schr\"{o}dinger equation, $H\Psi =\mathcal{E}_{n}\Psi,$ provides a linear second order differential equation with constant coefficients:
\begin{equation}
\frac{d^{2}\Psi }{d\phi ^{2}}-2i\beta \frac{d\Psi }{d\phi }+\xi \Psi =0,
\label{schr}
\end{equation}
where
\begin{equation}
\beta \equiv \frac{q\Phi _{\left( B\right) }+g\left\vert \mathbf{v}
\right\vert \Phi _{\left( E\right) }}{2\pi },\text{ \ }\xi \equiv
2mr_{0}^{2}\left( \mathcal{E}_{n}-qU\right) -\beta ^{2}. \label{beta1}
\end{equation}

Assuming that the eigenfunctions of Eq. (\ref{schr}) have the form,
\begin{equation}
\Psi=Ce^{i\ell\phi},\label{sol1}
\end{equation}
and replacing it in Eq. (\ref{schr}), one achieves the following characteristic equation for $\ell$:
\begin{equation}
\ell=\beta\pm\sqrt{\beta^{2}+\xi}.
\end{equation}

For the wave function $\Psi$ to be single-valued, the parameter $\ell$ must be an integer. In this case,
\begin{equation}
\beta \pm \sqrt{\beta ^{2}+\xi }=n. \label{cont}
\end{equation}
where $n=0,\pm 1,\pm 2,\ldots $With this condition, the system eigenenergies assume discrete values,
\begin{equation}
\mathcal{E}_{n}=\frac{1}{2mr_{0}^{2}}\left[ n-\frac{q\Phi _{\left( B\right)}+g\Phi _{\left( E\right) }\left\vert \mathbf{v}\right\vert }{2\pi }\right]^{2}+qU,\label{energy1}
\end{equation}
which depend on the magnetic/electric fluxes, and on the Lorentz-violating background, whose role can be elucidated as follows. In the absence of electromagnetic fields, the particle energy would be $\mathcal{E}_{n}=n/2mr_{0}^{2},$ the same value for a particle moving in the clockwise or counterclockwise sense, implying a circulation degenerescence.

If the electric flux is taken as null in Eq. (\ref{energy1}) (case without charge distribution along the solenoid), we are left only with the original AB result,
\begin{equation}
\mathcal{E}_{n}=\frac{1}{2mr_{0}^{2}}\left[ n-\frac{q\Phi _{\left( B\right) }
}{2\pi }\right]^{2}. \label{ab1}
\end{equation}
Note that this choice has eliminated the Lorentz-violating contribution, showing that the original AB effect can not be influenced by the Lorentz-violating term in this context. The result (\ref{ab1}) is the same one obtained in the presence only of magnetic field \cite{Ihn}, in which the dependence on $\Phi _{\left( B\right) }$ is enough for breaking the
circulation degenerescence. Indeed, it is easy to note that a positive $n$ yields lower energies (assuming $q$ is positive) than the corresponding negative values, $-\left\vert n\right\vert .$ This can be than interpreted as follows: the positive values of $n$ are compatible with a particle traveling in the same sense as the current in the solenoid, while the negative values of $n$ are compatible with a particle traveling in the opposite sense.

If now the magnetic flux is supposed to vanish, we retain only the AC
effect, whose eigenvalues,
\begin{equation}
\mathcal{E}_{n}=\frac{1}{2mr_{0}^{2}}\left[ n-\frac{g\Phi _{\left( E\right)
}\left\vert \mathbf{v}\right\vert }{2\pi }\right] ^{2}+qU,\label{acl}
\end{equation}
are clearly affected by the background magnitude, $\left\vert \mathbf{v}\right\vert$. Thus, we see that the LV background lifts the degenerescence even in the absence of the magnetic field (it occurs even for
a neutral particle deprived from magnetic moment). The degenerescence shift
is proportional to the background magnitude, in such a way the
Aharonov-Casher effect can be used as a tool to impose upper bounds on the
magnitude of the LV background, nonminimally coupled to fermions, as
stipulated in Eq. (\ref{cov}). Below, this is done taking as key-point the
geometrical AC phase.

As well-known, the induced phase is
\begin{equation}
\varphi =q\int \overset{\thicksim }{\mathbf{A}}\cdot \mathbf{d}\boldsymbol{\ell }\mathbf{,}  \label{phas}
\end{equation}
with the potential $\overset{\thicksim }{\mathbf{A}}$ defined\ as in Eq. (
\ref{A1}). The total induced phase,
\begin{equation}
\varphi =q\left( \Phi _{(B)}+\frac{g}{q}\Phi _{(E)}\left\vert \mathbf{v}
\right\vert \right) ,  \label{f1}
\end{equation}
obviously is a sum of the AB and AC phases, proportional to $\Phi _{\left(
B\right) }$ and $\Phi _{(E)}$, respectively. For a vanishing background, $
\left\vert \mathbf{v}\right\vert \rightarrow 0$, there remains only the
original AB phase. For a vanishing magnetic field, the attained AC phase is\
$g\Phi _{(E)}\left\vert \mathbf{v}\right\vert $. It is interesting to note
that the nonminimal coupling is able to induce an AC phase even for a
neutral particle without magnetic moment $\left( q=0,\mu =0\right) ,$ being
this gedanken situation used to impose an upper bound on the magnitude of
the product $g\left\vert \mathbf{v}\right\vert .$ The point is that, for
this particle, no usual physics may explain the induction of a topological
phase. Supposing an experimental ability to measure geometrical phases as
small as $10^{-4}$ rad \cite{cimmino}, we can affirm that the theoretical
AC phase induced for a neutral particle without magnetic moment $\left(
q=0,\mu =0\right)$ can not be larger than this value, that is,
\begin{equation}
2\pi r_{0}\left\vert \mathbf{E}\right\vert g\left\vert \mathbf{v}\right\vert
<10^{-4}\mathbf{\ rad}.  \label{condi}
\end{equation}
Taking $\left\vert \mathbf{E}\right\vert =10^{7}$ $Volt/m,$ $r_{0}=10^{-5}m$
(usual values of electric field and radius for 1D mesoscopic rings \cite{Nitta}), and working in the natural units system $\left( \hbar =c=1\right) ,
$ wherein $1Volt=11.7$ $eV,$ Eq. (\ref{condi}) leads to the following upper
bound:
\begin{equation}
g\left\vert \mathbf{v}\right\vert <10^{-8}\left( eV\right) ^{-1}.
\label{ub1}
\end{equation}

The table I below shows the results of $n$, $q$ and $\left\vert \mathbf{v}
\right\vert$ which provide discrete values for eigenenergy (\ref{energy1}).

\begin{widetext}
\begin{center}
\textbf{Table I}
\end{center}
\begin{equation*}
\begin{tabular}{clrc}
\hline
$\mathcal{E}_{n}=\frac{1}{2mr_{o}^{2}}\left[ n-\left( \frac{q\Phi
_{\left( B\right) }+g\Phi _{\left( E\right) }\left\vert \mathbf{v}
\right\vert }{2\pi }\right) \right] ^{2}+qU$\ \ \ \ \ \ \ \  & $
n=0,1,2,\ldots $ \ \ \ \ \ \ \ \  & $q>0$ \ \ \ \ \ \ \ \  & $\mathbf{v}
=\left\vert \mathbf{v}\right\vert $ \\
$\mathcal{E}_{n}=\frac{1}{2mr_{o}^{2}}\left[ n+\left( \frac{q\Phi
_{\left( B\right) }+g\Phi _{\left( E\right) }\left\vert \mathbf{v}
\right\vert }{2\pi }\right) \right] ^{2}-qU$\ \ \ \ \ \ \ \  & $
n=0,1,2,\ldots $ \ \ \ \ \ \ \ \  & $q<0$ \ \ \ \ \ \ \ \  & $\mathbf{v}
=-\left\vert \mathbf{v}\right\vert $ \\
$\mathcal{E}_{n}=\frac{1}{2mr_{o}^{2}}\left[ n-\left( \frac{q\Phi
_{\left( B\right) }+g\Phi _{\left( E\right) }\left\vert \mathbf{v}
\right\vert }{2\pi }\right) \right] ^{2}+qU$ \ \ \ \ \ \ \ \  & $
n=0,-1,-2,\ldots $ \ \ \ \ \ \ \ \  & $q>0$ \ \ \ \ \ \ \ \  & $\mathbf{v}
=\left\vert \mathbf{v}\right\vert $ \\
$\mathcal{E}_{n}=\frac{1}{2mr_{o}^{2}}\left[ n-\left( \frac{q\Phi
_{\left( B\right) }-g\Phi _{\left( E\right) }\left\vert \mathbf{v}
\right\vert }{2\pi }\right) \right] ^{2}-qU$\ \ \ \ \ \ \ \  & $
n=0,-1,-2,\ldots $\ \ \ \ \ \ \ \  & $q>0$\ \ \ \ \ \ \ \  & $\mathbf{v}
=-\left\vert \mathbf{v}\right\vert $ \\
$\mathcal{E}_{n}=\frac{1}{2mr_{o}^{2}}\left[ n-\left( \frac{q\Phi
_{\left( B\right) }-g\Phi _{\left( E\right) }\left\vert \mathbf{v}
\right\vert }{2\pi }\right) \right] ^{2}+qU$\ \ \ \ \ \ \ \  & $
n=0,1,2,\ldots $\ \ \ \ \ \ \ \  & $q>0$\ \ \ \ \ \ \ \  & $\mathbf{v}
=-\left\vert \mathbf{v}\right\vert $ \\
$\mathcal{E}_{n}=\frac{1}{2mr_{o}^{2}}\left[ n+\left( \frac{q\Phi
_{\left( B\right) }-g\Phi _{\left( E\right) }\left\vert \mathbf{v}
\right\vert }{2\pi }\right) \right] ^{2}-qU$ \ \ \ \ \ \ \ \  & $
n=0,1,2,\ldots $\ \ \ \ \ \ \ \  & $q<0$\ \ \ \ \ \ \ \  & $\mathbf{v}
=\left\vert \mathbf{v}\right\vert $ \\
$\mathcal{E}_{n}=\frac{1}{2mr_{o}^{2}}\left[ n-\left( \frac{q\Phi
_{\left( B\right) }-g\Phi _{\left( E\right) }\left\vert \mathbf{v}
\right\vert }{2\pi }\right) \right] ^{2}+qU$\ \ \ \ \ \ \ \  & $
n=0,-1,-2,\ldots $\ \ \ \ \ \ \ \  & $q>0$\ \ \ \ \ \ \ \  & $\mathbf{v}
=-\left\vert \mathbf{v}\right\vert $ \\
$\mathcal{E}_{n}=\frac{1}{2mr_{o}^{2}}\left[ n+\left( \frac{q\Phi
_{\left( B\right) }-g\Phi _{\left( E\right) }\left\vert \mathbf{v}
\right\vert }{2\pi }\right) \right] ^{2}-qU$\ \ \ \ \ \ \ \  & $
n=0,-1,-2,\ldots $\ \ \ \ \ \ \ \  & $q<0$\ \ \ \ \ \ \ \  & $\mathbf{v}
=\left\vert \mathbf{v}\right\vert $ \\ \hline
\end{tabular}
\ \ \ \ \
\end{equation*}
\end{widetext}

\section{The Aharonov-Bohm-Casher problem in the presence of a topological defect and Lorentz violation}\label{sec4}

In this section, we study the effect of a topological defect (disclination)
on the energy spectrum and wave function of a pointlike charge which
describes a circular path in a plane orthogonal to the\textbf{\ }magnetic
field, as in Sec. \ref{sec3}. We consider an infinitely long linear
disclination disposed along the z-axis, which is obtained by either removing
(positive-curvature disclination) or inserting (negative-curvature
disclination) a wedge of material \cite{furtado1}. If $\lambda $ is the
angle that defines the wedge, the metric of the disclination medium is
described by a non-Euclidean metric similar to Eq. (\ref{metrica}) \cite{katanaev}:
\begin{equation}
ds^{2}=dr^{2}+\alpha ^{2}r^{2}d\phi ^{2}+dz^{2},  \label{metrica2}
\end{equation}
where $\alpha =1+\lambda /2\pi $. This metric corresponds to a locally flat
medium with a conical singularity at the origin. The magnetic and electric
fields are modified by the disclination, being given as
\begin{equation}
\mathbf{B}=\frac{\Phi _{\left( B\right) }}{\alpha r}\delta \left( r\right)
\delta \left( \phi \right) \hat{z},\hspace{0.3cm}\mathbf{E}=\frac{\Phi _{(E)}}{2\pi \alpha
r}\hat{r},  \label{flux2}
\end{equation}
with the potential vector now written as $\mathbf{A}=(\Phi _{\left( B\right) }/2\pi
\alpha r)\hat{\phi}$. Following the procedure developed in Sec. \ref{sec3},
the time-independent Schr\"{o}dinger equation in the metric (\ref{metrica2})
is given by
\begin{widetext}
\begin{eqnarray}
-\frac{1}{2m}\frac{1}{\alpha ^{2}r_{0}^{2}}\left[\frac{d^{2}}{d\phi ^{2}}
-2i\left( \frac{q\Phi _{\left( B\right) }+g\left\vert \mathbf{v}\right\vert
\Phi _{\left( E\right) }}{2\pi }\right) \frac{d}{d\phi }-\left( \frac{q\Phi _{\left( B\right) }+g\left\vert \mathbf{v}\right\vert
\Phi _{\left( E\right) }}{2\pi }\right) ^{2}\right]\Psi \left( \phi \right)
+qU\Psi \left( \phi \right) =&\overset{\_\_}{\mathcal{E}}\Psi \left( \phi
\right). \label{sch2}
\end{eqnarray}
\end{widetext}The expression (\ref{sch2}) can be read as
\begin{equation}
\left[ \frac{d^{2}}{d\phi ^{2}}-2i\beta \frac{d}{d\phi }-\beta ^{2}\right]
\Psi \left( \phi \right) +\overset{\_\_}{\mathcal{E}}\Psi \left( \phi
\right) =0,  \label{diff2}
\end{equation}
where $\beta $ and $\varepsilon $ are given by Eq. (\ref{beta1}). The
discrete values for the eigenenergies, in accordance with Eqs. (\ref{sol1})
and (\ref{cont}), are given by
\begin{equation}
\overset{\_\_}{\mathcal{E}}_{n}=\frac{1}{2m\alpha ^{2}r_{0}^{2}}\left[ n-
\frac{q\Phi _{\left( B\right) }+g\Phi _{\left( E\right) }\left\vert \mathbf{v
}\right\vert }{2\pi }\right] ^{2}+qU.  \label{energy2}
\end{equation}
Here, we have the same result of the Eq. (\ref{energy1}). The only
difference is the presence of the parameter $\alpha $, which represents the
disclination. Despite this unique difference, it is physically very
important. The reason is that the topological defect affects the
eigenenergies, even with the particle not accessing the defect core region.
In the limit $\alpha \rightarrow 1$, when the curvature is zero, we recover
the results in Table I.

If we impose the electric flux as null in Eq. (\ref{energy2}), we obtain the
eigenenergies associated with the topological AB effect:
\begin{equation}
\overset{\_\_}{\mathcal{E}}_{n}=\frac{1}{2m\alpha ^{2}r_{0}^{2}}\left[ n-
\frac{q\Phi _{\left( B\right) }}{2\pi }\right] ^{2},  \label{abto}
\end{equation}
with the same meaning of Eq. (\ref{ab1}). Returning to Eq. (\ref{energy2}),
and taking $\Phi _{\left( B\right) }=0,$ we find the topological AC
eigenenergies:
\begin{equation}
\overset{\_\_}{\mathcal{E}}_{n}=\frac{1}{2m\alpha ^{2}r_{0}^{2}}\left[ n-
\frac{g\Phi _{\left( E\right) }\left\vert \mathbf{v}\right\vert }{2\pi }
\right] ^{2}+qU.  \label{topab}
\end{equation}
As in the foregoing section, the AC effect depends on the background, $\left\vert \mathbf{v}\right\vert,$ lifting the particle degenerescence in
the absence of external magnetic field, a result that holds even for a
neutral particle. In the limit of a vanishing curvature, $\alpha
\rightarrow 1$, Eqs. (\ref{abto})-(\ref{topab}) recover the AB and AC
eigenenergies, given by Eqs. (\ref{ab1})-( \ref{acl}). We should still
mention that the total induced geometrical phase for the topological ABC
problem remains given by Eq. (\ref{f1}), with $\Phi _{\left( B\right)
}=\alpha r_{0}\left\vert \mathbf{B}\right\vert ,$ $\Phi _{(E)}=2\pi \alpha
r_{0}\left\vert \mathbf{E}\right\vert $. With it, and in the absence of
magnetic feld, the upper bound (\ref{ub1}) is rewritten as
\begin{equation}
\alpha g\left\vert \mathbf{v}\right\vert <10^{-8}\mathbf{\ }\left( eV\right)^{-1}.
\end{equation}
\vspace{0.2mm}

\section{Topological Aharonov-Bohm-Casher in a two-dimensional harmonic
oscillator}

\label{sec5}

In this section, we study the dynamics of a charged particle in a space endowed with a topological defect (disclination), an ABC potential, and the LV nonminimal coupling, and interacting with a two-dimensional harmonic oscillator. A simpler version of this problem was addressed in Ref. \cite{furtado1} to study the quantum dynamics of a particle interacting harmonically with conical singularities in different physical contexts; in Ref. \cite{azevedo1} it were analyzed the changes introduced on the spectrum of a bound particle confined to move in a plane with a disclination with magnetic field. This type of system can be used to simulate the behavior of a charged particle in a continuous elastic medium with the topological defect and magnetic field \cite{furtado2}. Here, we reassess this kind of situation in the presence of the nonminimally coupled
Lorentz-violating background.

In this case, the metric is the same of Eq. (\ref{metrica2}). The Schr\"{o} dinger equation, using (\ref{H1}), becomes
\begin{equation}
\left[ \frac{1}{2m}\left( \mathbf{p}-q\overset{\thicksim }{\mathbf{A}}\right) ^{2}+qV+\frac{1}{2}m\omega ^{2}r^{2}\right] \Psi \left( r,\phi \right) =\overset{\thicksim }{\mathcal{E}}\Psi \left( r,\phi \right) .
\label{hosc}
\end{equation}
Inserting (\ref{ba}) and (\ref{ce}) into (\ref{hosc}), we arrive at
\begin{widetext}
\begin{align}
&  -\frac{1}{2m}\left[  \frac{\partial^{2}}{\partial r^{2}}+\frac{1}{r}\frac{\partial}{\partial r}+\frac{1}{\alpha^{2}r^{2}}\frac{\partial^{2}}{\partial\phi^{2}}-\left(2iq\frac{\Phi_{\left(  B\right)}
}{2\pi\alpha r}+2ig\frac{\left\vert \mathbf{v}\right\vert \Phi_{(E)}}{2\pi\alpha r}\right)  \frac{1}{\alpha r}\frac{\partial} {\partial\phi}-2qg\frac{\left\vert \mathbf{v}\right\vert
\Phi_{\left(  E\right)}\Phi_{\left(B\right)}}{\left(2\pi\alpha r\right)  ^{2}}-\left(  \frac{\Phi_{\left(  B\right)  }}{2\pi\alpha r}\right) ^{2}\right]  \times\nonumber\label{serie}\\[-2mm]& \\[-2mm]
&  \times\Psi\left(  r,\phi\right)  +\left(qU+\frac{1}{2}m\omega^{2}
r^{2}\right)  \Psi\left(  r,\phi\right)  =\overset{\thicksim}{\mathcal{E}}
\Psi\left(  r,\phi\right)  .\nonumber
\end{align}
\end{widetext}

Let us assume the eigenfunction of the form
\begin{equation}
\Psi \left( r,\phi \right) =e^{i\ell \phi }R\left( r\right) ,
\end{equation}
which satisfies the usual asymptotic requirements and finiteness at the
origin for a bound state. This leads to the radial equation
\begin{equation}
\left( \frac{d^{2}}{dr^{2}}+\frac{1}{r}\frac{d}{dr}-\frac{l^{2}}{\alpha
^{2}r^{2}}\right) R\left( r\right) -\gamma ^{2}r^{2}R\left( r\right)
-k^{2}R\left( r\right) =0,  \label{radial}
\end{equation}
where $l=\ell -q\beta \Phi _{\left( B\right)}-g\zeta \left\vert \mathbf{v}
\right\vert \Phi _{(E)},$ $\gamma =m\omega ,$ $k^{2}=2m\left( qU-\overset{
\thicksim }{\mathcal{E}}\right) $ and $\zeta =1/2\pi $. The radial Eq\textbf{
.} (\ref{radial}) is well known and its solution is the degenerated
hypergeometric function
\begin{equation}
R\left( r\right) =F\left( \eta ,\frac{\alpha +\left\vert l\right\vert }{
\alpha };\gamma r^{2}\right) ,  \label{serieh}
\end{equation}
with $\eta =\left[ \alpha \left( 2\gamma +k^{2}\right) +2\left\vert
l\right\vert \gamma \right] /4\gamma \alpha $. Moreover, if $\eta $ is
equal to $0$ or a negative integer, the series terminates and the
hypergeometric function becomes a polynomial of degree $n$. This condition
guarantees that the hypergeometric equation has a regular singularity in the
origin, which is essential for our treatment of the physical system
considered since we have sources located in the origin. Therefore, the
series in Eq. (\ref{serieh}) must converge if we consider that
\begin{equation}
\eta =-n.\label{cond}
\end{equation}
This condition also guarantees the normalization of the wavefunction. Using
Eq. (\ref{cond}), we obtain the discrete eigenenergies as below
\begin{eqnarray}
\overset{\thicksim }{\mathcal{E}}_{n,\ell } &=&\omega \left( 2n+\frac{1}{
\alpha }\left\vert \ell -q\zeta \Phi _{\left( B\right) }-g\zeta \left\vert
\mathbf{v}\right\vert \Phi _{(E)}\right\vert +1\right) +  \notag \\[-0.2mm]
&& \\[-0.2mm]
&+&qU,\notag  \label{energyo}
\end{eqnarray}
where $n=0,1,2,\ldots $. We see that the presence of the parameter $\alpha $
, the magnetic flux, and the LV background, are able to break the degeneracy
of the energy levels. In the absence of background field $\mathbf{v}$
reproduces a result already known in \cite{Phase1}.The energy eigenfunction
is given by
\begin{align}
\Psi \left(r,\phi \right) & =\allowbreak C_{n\ell }\gamma ^{\frac{1}{2}
\left(\frac{\alpha +\left\vert l\right\vert }{\alpha }\right) }r^{\frac{
\left\vert l\right\vert }{\alpha }}e^{-\frac{1}{2}\gamma r^{2}}e^{i\ell \phi
}\times  \notag  \label{efe} \\[-2mm]
& \\[-2mm]
& \times _{1}F_{1}\left(-n,\frac{\alpha +\left\vert l\right\vert }{\alpha}
;\gamma r^{2}\right),  \notag
\end{align}
where $C_{n\ell }$ is a normalization constant. We note that the
eigenfunctions are modified in a space with a disclination, even that the
particle does not have access to the defect region. In the limit $\alpha
\rightarrow 1$, we obtain the harmonic oscillator eigenenergies modified by
the AB and AC contributions.

Let us analyze the result of Eq. (\ref{energyo}) in the absence of the
electric flux, for which the topological AB harmonic oscillator eigenergies
are
\begin{equation}
\overset{\thicksim }{\mathcal{E}}_{n,\ell }=\omega \left( 2n+\frac{1}{\alpha
}\left\vert \ell -q\beta \Phi _{\left( B\right) }\right\vert +1\right) .
\label{energyo2}
\end{equation}
Note that the energy depends only on the magnetic flux and on the
disclination. The ground state of (\ref{energyo2}) is given by
\begin{equation}
\overset{\thicksim }{\mathcal{E}}_{0,0}=\omega \left( 1-\frac{\left\vert
q\right\vert \zeta \Phi _{\left( B\right) }}{\alpha }\right) .
\end{equation}
We can see that the ground state is also affected by disclination. As a
final analysis, we can obtain the conventional harmonic oscillator making $
\Phi_{\left( B\right) }\rightarrow 0,\Phi_{\left( E\right) }\rightarrow 0$
and $\alpha \rightarrow 1,$ simultaneously.

Now, we examine the result of Eq. (\ref{energyo}) in the absence of magnetic
flux, for which the harmonic oscillator eigenenergies are
\begin{equation}
\overset{\thicksim }{\mathcal{E}}_{n,\ell }=\omega \left( 2n+\frac{1}{\alpha
}\left\vert \ell -g\zeta \left\vert \mathbf{v}\right\vert \Phi
_{(E)}\right\vert +1\right) +qU,\label{acpo}
\end{equation}%
which depend on the parameter $\alpha $, on the electric flux and the
background magnitude, $\left\vert \mathbf{v}\right\vert $. Similarly to the
previous cases, we see that LV background is able to modify the harmonic
oscillator eigenenergies even for a chargeless particle without magnetic
moment $\left( q=0,\mu =0\right)$. The corresponding ground state energy
for Eq. (\ref{acpo}), $n=\ell =0$, is
\begin{equation}
\overset{\thicksim }{\mathcal{E}}_{0,0}=\omega \left( 1-\frac{g\zeta
\left\vert \mathbf{v}\right\vert \Phi_{\left( E\right) }}{\alpha }\right)
+qU.
\end{equation}

\section{Conclusions}

\label{sec6}

The ABC problem for a particle moving in a mesoscopic circular path was
analyzed (a) in the presence of a LV background nonminimally coupled to the
fermion and gauge fields as proposed in Ref. \cite{Phase1}, (b) in a space
endowed with a topological defect (disclination) and considering the LV
nonminimal coupling, (c) and for a particle in a continuous elastic medium
with topological defect and a LV background.

The background field vector $v$ is chosen in such a way that it contributes
to the topological phase. For the case (a), we have shown that the discrete
eigenenergies for the circling particle carry contributions coming from the
magnetic and electric fluxes. We notice that the nonminimal coupling is able
to lift the circulation degenerescence in the absence of magnetic field even
for chargeless particle without magnetic moment $\left( \mu =0\right)$. The
induced AC geometrical phase is used to impose a good upper bound, $%
g\left\vert \mathbf{v}\right\vert <10^{-8}\ \left( eV\right) ^{-1},$ by
means of a different route from the one of Ref. \cite{Nonmini}.

In the case (b), the particle eigenenergies obey the same pattern of case
(a), except for the inclusion of the parameter $\alpha$ that characterizes
the topological defect. The eigenenergies are modified by the topology of
the space, even noting that the particle does not have access to the defect
region. In the limit $\alpha\rightarrow1$ we recover the results of the case
(a). In the case (c), we verify that the energy spectrum and wave functions
are modified by the presence of a disclination. It is important to remark
that the region outside the defect has zero curvature. We have found that
the ground state also is affected by the disclination. Finally, we notice
that the topology plays a role as similar to the magnetic field along the
axis of the AB problem, once both do not exist in the region of space where
the particle moves.

\begin{center}
{\small {\textbf{ACKNOWLEDGMENTS}} }
\end{center}

M. T. D. Orlando thanks CNPq, H. Belich and E. O. Silva thank the Brazilian
Ministry of Science and Technology for the financial support. M. M. Ferreira
Jr is grateful to FAPEMA, CAPES and CNPq (Brazilian agencies) for financial
support.

\end{document}